\documentclass[11pt,twoside]{article}
\usepackage[pdftex]{graphicx}
\usepackage{amsmath}
\usepackage{amssymb}
\usepackage{cite}

\begin{document}
\newcommand{\pst}{\hspace*{1.5em}}

\newcommand{\rigmark}{\em Journal of Russian Laser Research}
\newcommand{\lemark}{\em Volume 32, Number 5, 2011}

\newcommand{\be}{\begin{equation}}
\newcommand{\ee}{\end{equation}}
\newcommand{\bm}{\boldmath}
\newcommand{\ds}{\displaystyle}
\newcommand{\bea}{\begin{eqnarray}}
\newcommand{\eea}{\end{eqnarray}}
\newcommand{\ba}{\begin{array}}
\newcommand{\ea}{\end{array}}
\newcommand{\arcsinh}{\mathop{\rm arcsinh}\nolimits}
\newcommand{\arctanh}{\mathop{\rm arctanh}\nolimits}
\newcommand{\bc}{\begin{center}}
\newcommand{\ec}{\end{center}}

\thispagestyle{plain}

\label{sh}


\begin{center} {\Large \bf
SU(2) INVARIANTS OF SYMMETRIC QUBIT STATES}
\end{center}

\bigskip

\bigskip

\begin{center} {\bf
Swarnamala Sirsi$^1$  and Veena Adiga$^{1,2,*}$  
}\end{center}

\medskip

\begin{center}
{\it
$^1$Department of Physics, Yuvaraja's College\\
University of Mysore, Mysore-05, India

\smallskip

$^2$St.Joseph's College (autonomous), Bengaluru-27 , India 
}
\smallskip

$^*$Corresponding author e-mail:~~~vadiga11@gmail.com\\
\end{center}

\begin{abstract}\noindent
Density matrix for N-qubit symmetric state or spin-j  state (j = N/2) is expressed in terms of the well known Fano statistical tensor parameters.
Employing the multiaxial representation \cite {Ravishankar}, wherein a spin-j density matrix is shown to be characterized by j(2j+1) axes and 2j 
real scalars, we enumerate the number of invariants constructed out of these axes and scalars. These invariants are explicitly calculated in the particular case 
of pure as well as mixed spin-1 state.
\end{abstract}

\medskip

\noindent{\bf Keywords:}
SU(2)Invariants, symmetric state, density matrix, quantum entanglement. 

\section{Introduction}
\pst
The problem of enumeration of local invariants of quantum state described by a density matrix $\rho$ is important in the context of quantum 
entanglement. Nonlocal correlations in quantum systems reflect entanglement between its parts. Genuine non- local properties should be described in 
a form invariant under local unitary operations. Two N-qubit states are said to be locally equivalent if one can be transformed into the other by 
local operations. i.e., ${\rho^{\prime}}$ = $U\rho U^{\dagger}$   where $U \in SU(2)^{\times{N} }$ and  the two quantum states  $\rho$ and 
${\rho^{\prime}}$ are said to be equally entangled. \\

A general prescription to identify the invariants associated with a multiparticle system have been outlined by Linden et al., \cite{Linden}. 
Well known algebraic methods for generating invariants  already exists in literature \cite{{Osterloh}, {Barnum}, {Carteret},
{Grassl}}. Williamson et al.,\cite{Williamson}  have presented a geometric approach 
for constructing SU(2) and SL(2,C) invariants. Makhlin \cite{Makhlin} has presented a complete set of 18 local polynomial invariants of two qubit 
mixed states and demonstrated the usefulness of these invariants to study entanglement. As the number of subsystems increases, the problem of 
identifying and interpreting independent invariants rapidly becomes very complicated. Usha et al., \cite{Usha} have shown that a set of 6 invariants which is a subset of a more general set of 18 invariants proposed by Makhlin \cite{Makhlin}
is sufficient to characterize the non local properties of a symmetric two qubit system. We also focus on symmetric two qubit states as the problem 
of identifying independent invariants become easier. Our approach makes use of the geometrical multiaxial representation of an arbitrary spin-j 
density matrix \cite{Ravishankar} which is completely characterized by a set of j(2j+1) axes and 2j real positive scalars.\\     

The paper is organized as follows: 
In Sec. 2,\,\, the decomposition of a density matrix in terms of well known Fano statistical tensor parameters is presented. Discussion 
of multiaxial description of the density matrix using Wigner-D matrices is outlined and the invariants associated with N-qubit symmetric state
are enumerated in section 3. In Sec.~4,\,\,  We have explicitly calculated the invariants of two qubit symmetric mixed as well as the most general 
pure state. To make our task easier, we have considered the Special Lakin Frame which is widely used in nuclear reactions. 
\section{Symmetric subspace}
\pst
Here we are interested in the set of N-particle pure states that remain unchanged by permutations of individual 
particles. Symmetric states offer elegant mathematical analysis as the dimension of the Hilbert space reduces drastically from $2^{N}$ to (N + 1), 
when  N qubits respect exchange symmetry. Such a Hilbert space is considered to be spanned by the eigen states $\{| j,m\rangle;-j\leq m\leq+j\}$ of 
angular momentum operators $J^{2}$ and $J_{z}$ , where $j=\frac{N}{2}$. Analyzing a general state of N-particle spin-1/2 system represented 
by  the density matrix of dimension $2^{N}\,\times\,2^{N}$ is difficult because the system's Hilbert space increases exponentially with the number 
of qubits N. Fortunately, a large number of experimentally relevant states possesses symmetry under particle exchange and this property allows us to
significantly reduce the computational complexity. Completely symmetric systems are experimentally interesting, largely because it is often easier to
nonselectively address an entire ensemble of particles rather than individually address each member and it is possible to express the dynamics of 
these systems using only symmetry preserving operators. The symmetric subspace therefore provides a convenient, computationally accessible class 
of spin states. Specifically, if we have N two level atoms, each atom may be represented as a spin- 1/2 system and theoretical analysis can be 
carried out in terms of collective spin operator $\vec {J}$ = $\frac {1}{2}{\Sigma^{N} _{\alpha=1}}\vec {\sigma}_{\alpha}$. Here $\vec{\sigma}_{\alpha}$ denote
the Pauli spin operator of the $\alpha$th qubit. 
The standard expression for the most general spin-j density matrix in terms of Fano statistical tensor parameters $t^{k}_{q}\,{'}s$ 
is given by 
\begin{equation}
\rho(\vec{J}) = \frac {Tr(\rho)}{(2j+1)}\sum^{2j}_{k=0}\,\sum^{+k}_{q=-k}\,\, t^{k}_{q}\, \tau^{k^{\dagger}}_{q}(\vec{J})\,\,,
\end{equation} 
where   $\tau^{k}_{q}$ \, (with $\tau^{0}_{0} = I$ ,the identity operator) are irreducible tensor operators of rank `k' in the 2j+1 dimensional 
spin space with projection `q' along the axis of quantization in the real 3-dimensional space. The $\tau^{k}_{q}$ satisfy the orthogonality 
relations
\begin{equation}
Tr({\tau^{k^{\dagger}}_{q}\tau^{k^{'}}_{q^{'}}})= (2j+1)\,\delta_{kk^{'}} \delta_{qq^{'}}\,.
\end{equation}
Here the normalization has been chosen so as to be in agreement with Madison convention \cite {Satchler}. The spherical tensor parameters $t^{k}_{q}\,{'}s$
which characterize the given system are the average expectation values given by 
$t^{k}_{q} = \frac {{Tr({\rho\,\tau^{k}_{q}})}}{Tr \rho}$.
Since $\rho$ is Hermitian and $\tau^{k^{\dagger}}_{q} = (-1)^{q}\tau^{k}_{-q}$,  \,\, $t^{k}_{q}$ satisfy the condition 
\begin{equation}
t^{k^{*}}_{q} = (-1)^{q}\,t^{k}_{-q}\,\,.
\end{equation}
The spherical tensor parameters ${t^{k}_{q}}\,{'}s$ have simple transformation properties under co-ordinate rotation in the 3-dimensional space. In the 
rotated frame ${t^{k}_{q}}\,{'}s$ are given by 
\begin{equation}
(t^{k}_{q})^{R} = \sum^{+k}_{q^{'}=-k}\,\, D^{k}_{q^{'}q}(\phi,\theta,\psi)\,t^{k}_{q^{'}}\,\,,
\end{equation}
where $D^{k}_{q^{'}q}(\phi,\theta,\psi)$ denote Wigner-D rotation matrix described by Euler angles $(\phi, \theta, \psi)$.

\section{Multiaxial description of density matrix}
\pst
It has already been shown \cite{Ravishankar} that a spin-j density matrix is characterized by j(2j+1) axes and 2j real positive scalars. 
For the sake of completeness, we reproduce it here; 
In general $t^{k}_{\pm k}$ can be made zero for any k by suitable rotation. i.e.,
\begin{equation}
(t^{k}_{\pm k})^{R}=0=\sum _{q^{\prime}=-k}^{+k}D^{k}_{q^{\prime},\pm k}{(\phi,\theta,\psi)}t^{k}_{q^{\prime}}\,. 
\end{equation}
 
Using the well known Wigner expression for the rotation matrix $D^{k}$, the above equation can be written as 
\begin{equation}
(t^{k}_{\pm k})^{R}=0=[\pm^{sin}_{cos}(\theta/2)]^{2k}\,exp[i(\phi+\psi)]\sum _{r=0}^{2k}C_{r}Z^{r}\,\,,
\end{equation}
where the complex variable $Z=cot(\theta/2)e^{-i\phi}$\, in the case of $(t^{k}_{+k})^{R}=0$ and $Z=tan(\theta/2)e^{-i(\phi+\pi)}$\, in the case of 
$(t^{k}_{-k})^{R}=0$.
The expansion coefficients $C_{r}$ in the polynomial are the same in both the cases  and is given by $C_{r}= 
\left(
\begin{array}{c}
2k\\
k+q
\end{array}\right)^{\frac{1}{2}}t^{k}_{q}\;=\left(\begin{array}{c}
2k\\
r
\end{array}\right)^{\frac{1}{2}}t^{k}_{r-k}\;.$
By solving the above polynomial equation, one can get in general two sets of k-coordinate frames in which $(t^{k}_{\pm k})=0$. Explicitly, 
if $t^{k}_{k}$= 0 in a coordinate system where $\hat{Z}$ axis is directed along $(\theta,\phi)$ in the laboratory,
$t^{k}_{-k}$ = 0 in a coordinate system where $\hat{Z}$ axis is directed along $(\pi-\theta, \phi+\pi)$. One set is obtained by the other by 
inverting the $\hat{Z}$-axis. Therefore it is sufficient to enumerate the k independent solutions $\hat{Q}_{i}(\theta_{i}, \phi_{i})$, \,\, i=1,2,....k 
which constitute any arbitrary $t^{k}_{q}$ as a spherical tensor product of the form
\begin{equation}
t^{k}_{q} = r_{k}(...((\hat{Q}_{1}\otimes\hat{Q}_{2})^{2}\otimes\hat{Q}_{3})^{3}\otimes....)^{k-1}\otimes\hat{Q}_{k})^{k}_{q} \,, 
\end{equation}
where \begin{equation}
(\hat{Q}_{1}\otimes\hat{Q}_{2})^{2}_{q} = \sum _{q_{1}}C(11k;q_{1}q_{2}q)(\hat{Q}_{1})_{q_{1}} (\hat{Q}_{2})_{q_{2}}\,\,; \, \,\,\,
(\hat{Q})_{q}= \sqrt{\frac{4\pi}{3}}\,\,Y_{1q}(\theta,\phi). 
\end{equation}
Here $C(11k;q_{1}q_{2}q)$ is the Clebsch Gordan Co-efficient and $Y_{1q}(\theta,\phi)$ are the well known spherical harmonics. If one of the 
$\hat{Q}_{i}\,{'}s$ is inverted, sign of equation(7) is changed.  
Hence it is possible to choose k axes $\hat{Q}_{i}{'}s$ , \,\, i=1,2,....k in such a way that $r_{k}$ is always positive. Each axis requires two independent 
parameters $(\theta, \phi)$  to characterize it, hence the k axes together with the overall multiplicative factor account for exactly (2k+1) real 
parameters needed to specify a spherical tensor $t^{k}_{q}$ satisfying equation (3). Thus any spherical tensor of rank k can be represented
geometrically by a set of k vectors $\hat{Q}_{i}$ on the surface of a sphere of radius r. Consequently,
the state of a spin-j assembly can be represented geometrically by a set of 2j spheres, one corresponding to each value of k, k=1....2j, the kth 
sphere having k vectors specified on its surface.     
    
Since $(\hat{Q}_{i}(\theta_{i},\phi_{i})\otimes \hat{Q}_{j}(\theta_{j},\phi_{j}))^{0}_{0}$ \,\, is an invariant ($i\neq j$), one can construct 
in general $\left(\begin{array}{c}
j(2j+1)\\
2
\end{array}\right)$ invariants from j(2j+1) axes. Together with 2j real positive scalars, there are $\left(\begin{array}{c}
j(2j+1)\\
2
\end{array}\right)$ + 2j invariants characterizing spin-j density matrix. Thus using this multiaxial parametrization of density matrix, we enumerate the total number of SU(2) invariants characterizing a 
spin-j density matrix . Let us consider the example of two qubit symmetric state for a detailed discussion. 
\section{Invariants of two qubit symmetric state or spin-1 state}
\subsection{Pure spin-1 state}
\pst
Consider the  direct product $|\psi_{1}\rangle\otimes|\psi_{2}\rangle$ of two spinors in the qubit basis as 
\begin{eqnarray}|\psi _{12}\rangle =  \left(\begin{array}{cc}
 cos\frac {\theta_{1}}{2} \cr
 sin\frac {\theta_{1}}{2}e^{i\phi_{1}}\cr
 \end{array}\right)\otimes\left(\begin{array}{cc}
 cos\frac {\theta_{2}}{2} \cr
 sin\frac {\theta_{2}}{2}e^{i\phi_{2}}\cr
 \end{array}\right)                  
= \left(\begin{array}{cc}
  cos\frac {\theta_1}{2}cos\frac {\theta_2}{2} \cr
 cos\frac {\theta_1}{2}sin\frac {\theta_2}{2}e^{i\phi_{2}} \cr
 sin\frac {\theta_1}{2}cos\frac {\theta_2}{2}e^{i\phi_{1}}\cr
 sin\frac {\theta_1}{2}sin\frac {\theta_2}{2}e^{i(\phi_{1}+\phi_{2})}  \cr
 \end{array}\right),\end{eqnarray} 
 $0\leq\theta_{1,2}\leq\pi$\,, $0\leq\phi_{1,2}\leq 2\pi$\,.  In the symmetric angular momentum subspace $|11\rangle$, $|10\rangle$,
 $|1-1\rangle$, the combined state will have the form 
 \begin{equation}
  |\psi _{12}\rangle_{sym}= \left(\begin{array}{cc}
 cos\frac {\theta_1}{2}cos\frac {\theta_2}{2} \cr
 \frac{1}{\sqrt 2}(cos\frac {\theta_1}{2}sin\frac {\theta_2}{2}e^{i\phi_{2}}+ sin\frac {\theta_1}{2}cos\frac {\theta_2}{2}e^{i\phi_{1}})\cr
 sin\frac {\theta_1}{2}sin\frac {\theta_2}{2}e^{i(\phi_{1}+\phi_{2})}\cr
 \end{array}\right).\end{equation}
 
Since the two directions  $(\theta_{1},\phi_{1})$, $(\theta_{2},\phi_{2})$ associated with the above two spinors define a plane, we choose this
to be the xz-plane with respect to a frame $x_{0}y_{0}z_{0}$ with $\hat{z}_{0}$  being the bisector of the above two directions. Thus the azimuths 
of the above two directions $(\theta_{1},\phi_{1})$, $(\theta_{2},\phi_{2})$ with respect to $x_{0}$ are respectively 0 and $\pi$. If the angular 
separation between the two directions is $2\theta$, then the state $|\psi\rangle$ has the explicit form 
 \begin{equation}
|\psi\rangle = \frac {\sqrt{2}}{\sqrt{1+cos^{2}\theta}}[cos^{2}\frac{\theta}{2}|11\rangle_{\hat{Z}_0}-sin^{2}\frac{\theta}{2}|1-1\rangle_{\hat{Z}_0}].
\end{equation}
The density matrix corresponding  to the above state is given by
\begin{equation}
{\rho_s}=\frac{2}{(1+cos^{2}\theta)}
\left(\begin{array}{cccc}
cos^{4}\frac{\theta}{2} & 0 & -sin^{2}\frac{\theta}{2}cos^{2}\frac{\theta}{2}   \cr

 0 & 0 & 0     \cr

-sin^{2}\frac{\theta}{2}cos^{2}\frac{\theta}{2} & 0 & sin^{4}\frac{\theta}{2}  \cr
\end{array}\right).
\end{equation} 
Comparing equation (12) with the standard representation of the density matrix  
\begin{equation}
{\rho_s}=\frac{Tr(\rho)}{3}
\left (\begin{matrix}
1+\sqrt {\frac{3}{2}}\,t^{1}_{0}+ \frac {t^{2}_{0}}{\sqrt{2}} &&&  \sqrt\frac{3}{2}\,(t^{1}_{-1}+t^{2}_{-1}) &&& \sqrt{3}\,t^{2}_{-2}  \cr

 -\sqrt\frac{3}{2}\,(t^{1}_{1}+t^{2}_{1}) &&& 1-\sqrt{2}\,t^{2}_{0} &&& \sqrt\frac{3}{2}\,(t^{1}_{-1}-t^{2}_{-1})   \cr

\sqrt{3}\,t^{2}_{2} &&& -\sqrt\frac{3}{2}\,(t^{1}_{1}-t^{2}_{1}) &&&   1-\sqrt {\frac{3}{2}}\, t^{1}_{0}+ \frac {t^{2}_{0}}{\sqrt{2}}  \cr
\end{matrix}\right),\end{equation}
we get the non-zero  $t^{k}_{q}\,{'}s$  to be
\[t^{1}_{0} = \frac{\sqrt{6}cos\theta}{1+cos^{2}\theta} \,\,, \,\,t^{2}_{0} = \frac{1}{\sqrt 2}\,\,,\,\,
t^{2}_{2} = t^{2}_{-2} = \frac {\sqrt{3}sin^{2}\theta}{2(1+cos^{2}\theta)}\,\,.\]
Since $t^{1}_{\pm 1} = 0$\,, $\hat{z}_{0}$  itself is the axis ($\hat {Q}_{1}$) associated with $t^{1}$. As $t^{1}_{0}= r_{1}(\hat{Q}_{1})^{1}_{0}$\,, 
\begin{equation}
r_{1} = \frac {t^{1}_{0}}{(\hat{Q}_{1})^{1}_{0}}\,\,\,.
\end{equation}
Solving for  the polynomial equation (6) for $t^{2}$, we get $(\hat{Q}_{2})^{1}_{q} = \sqrt{\frac{4\pi}{3}}Y^{1}_{q}(\theta, 0)$ and  
$(\hat{Q}_{3})^{1}_{q} = \sqrt{\frac{4\pi}{3}}Y^{1}_{q}(\theta, \pi)$. 
Hence 
\begin{equation}
r_{2} =  \frac {t^{2}_{0}}{(\hat{Q}_{2}\otimes \hat{Q}_{3})^{2}_{0}} = \frac {t^{2}_{2}}{(\hat{Q}_{2}\otimes \hat{Q}_{3})^{2}_{2}}\,\,.
\end{equation}
The invariants associated with the most general pure spin-1 state are
\begin{equation} I_{1} = r_{1}\,\,,\, I_{2} = r_{2}\,\,,\, I_{3}= (\hat{Q}_{1}\otimes\hat{Q}_{2})^{0}_{0}\, \,, I_{4}= (\hat{Q}_{1}\otimes\hat{Q}_{3})^{0}_{0}\,\,\,, 
\,\, I_{5} = (\hat{Q}_{2}\otimes \hat{Q}_{3})^{0}_{0}\,\,.\end{equation}                                                                
Explicitly,   
\begin{equation}
I_{1} = \frac{\sqrt{6}|cos\theta|}{1+cos^{2}\theta}\,\,, \,I_{2} = \frac{\sqrt{3}}{1+cos^{2}\theta}\,\,, 
\,I_{3} = I_{4}= -\frac {cos\theta}{\sqrt 3} \,\,,\, I_{5} = -\frac {cos2\theta}{\sqrt 3}  .\end{equation}
It is clear from equation (11) that the state $|\psi\rangle$ is separable  for $\theta = 0$ and $\pi$ . Hence the invariants in the case of pure 
spin-1 separable states are 
\[ I_{1} = \sqrt{\frac{3}{2}}\,\,,\, I_{2} = \frac{\sqrt 3}{2}\,\,,\, I_{3} = I_{4} = \mp \frac{1}{\sqrt 3}\,\,,\, I_{5} = -\frac{1}{\sqrt 3} \,\,.\]

\subsection{Mixed spin-1 state}
\pst
Consider the example of channel spin-1 system which plays an important role in nuclear physics experiments like hadron scattering and reaction 
processess \cite{{Raichle},{Meyer},{H.O},{Thorngren},{Engblom}}. A beam of nucleons colliding with a proton target provides such an example. If both the beam 
and the target are prepared to  be in mixed states, then the corresponding density matrices  are given by
\begin{equation}
\rho(i) = \frac{1}{2}\,[\, I + \vec\sigma (i)\cdot\vec p(i)\,] = \frac {1}{2}\,\sum_{k,q}\,t^{k}_{q}(i)\,\tau^{k^{\dagger}}_{q}(i);\,\, \, i=1,2.
\end{equation}
where ${\vec p(i)}$ are the polarization vectors and $\vec\sigma(i)'$s are the Pauli spin matrices.

The combined density matrix is the direct product of the individual density matrices  
\begin {equation}
\rho_{c} = \rho(1) \otimes \rho(2)\,\,.
\end{equation}

Eventhough the combined density matrix is a direct product of individual matrices, in this case, entanglement appears due to the projection of 
the combined density matrix onto the desired spin-1 space. 
While solving this problem the Special Lakin frame (SLF) which is widely used in studying nuclear reactions is considered : Choose $\hat z_0$ to be
along $\vec p(1)+\vec p(2)$. Since $\vec p(1)$, $\vec p(2)$ together define a plane in any general situation, we choose ${\hat x_0}$\, to be in this
plane such that the azimuths of ${\vec p(1)}$ , $\vec p(2)$ with respect to  $\hat x_0$  are respectively 0 and ${\pi}$. The $\hat y_0$ axis is then 
chosen to be along ${\hat z_0\times \hat x_0}$. The frame so chosen is indeed the special Lakin frame (SLF)\, as here $t^{1}_{\pm 1} = 0$ and $t^{2}_{2} = t^{2}_{-2}$ .
Choose a simple case of $|\vec p(1)| = |\vec p(2)|= p $, then we get $t^{2}_{\pm 1} = 0 $ in SLF.
\begin{figure}[h]
\bc \includegraphics[width=8.6cm]{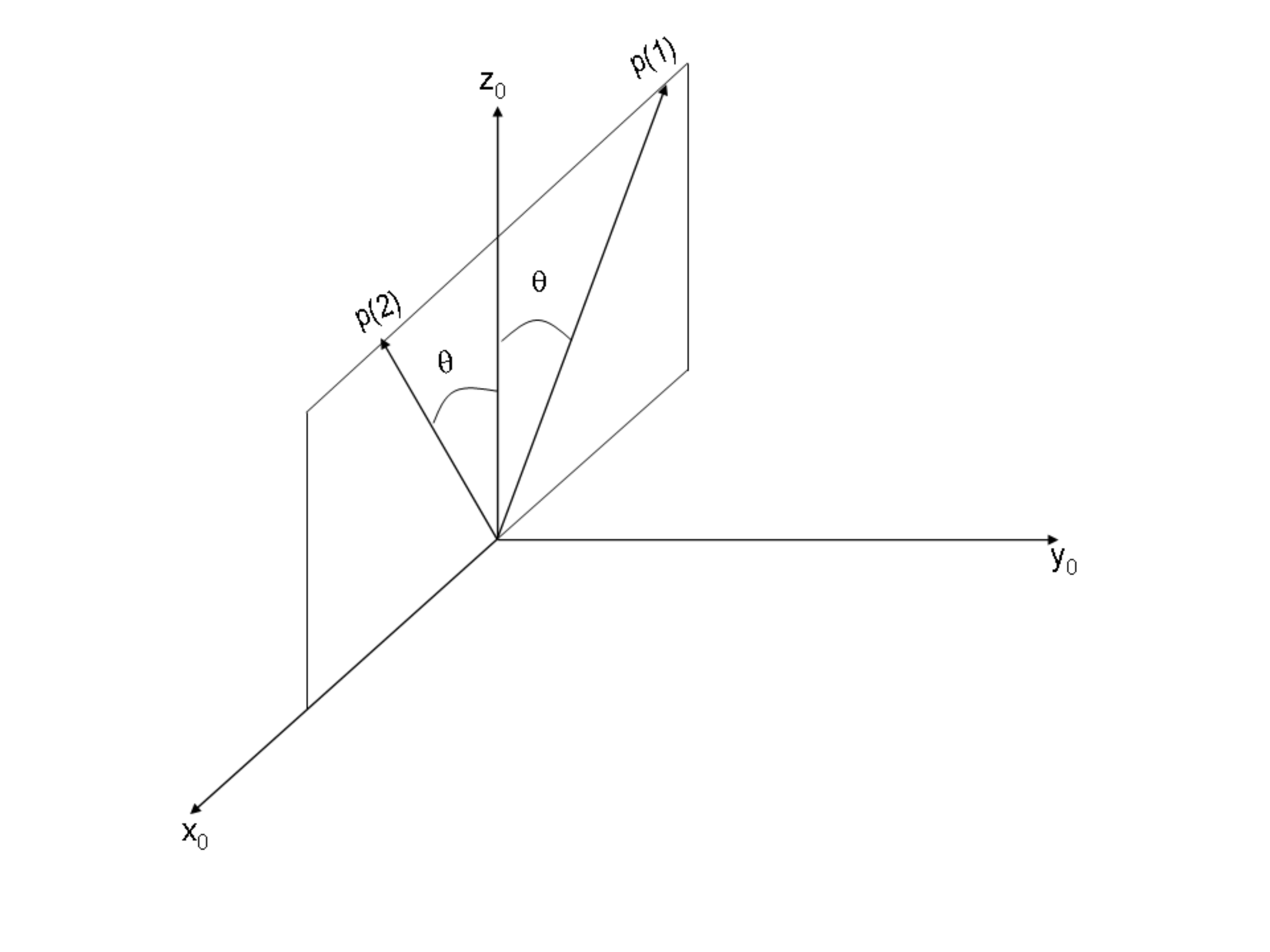}
\ec
\vspace{-4mm}
\caption{
$x_0y_0z_0$\, frame with mean spin direction $\hat z_0$ as the bisector of two directions ${\vec{p}(1)}$ and $\vec{p}(2)$}.
\end{figure}
The density matrix so obtained for spin-1 mixed system in the symmetric subspace $|11\rangle$, $|10\rangle$ and $|1-1\rangle$ is
\begin{equation}
{\rho_s}=\frac{1}{(3+p^{2}cos2\theta)}
\left (\begin{matrix}
(1+pcos\theta)^{2} &&& 0 &&& -p^{2}sin^{2}\theta\\

\\ 0 &&& 1-p^{2} &&& 0 \\

\\ -p^{2}sin^{2}\theta &&& 0 &&& (1-pcos\theta)^{2}
\end{matrix}\right)\,.\\
\end{equation} \\ 
Observe that when p=1, the mixed state density matrix is exactly the same as that of pure state density matrix as given by equation (12).  
Comparing the above density matrix with the standard form (equation (13)), we get the non-zero $t^{k}_{q}\,'s$ as
\[t^{1}_{0} = \frac{2\sqrt{6}pcos\theta}{(3+p^{2}cos2\theta)}\,\,,\, 
t^{2}_{0} = \frac{\sqrt{2}p^{2}(1+cos^{2}\theta)}{(3+p^{2}cos2\theta)}\,\,,\,t^{2}_{2} = \frac{\sqrt{3}p^{2}sin^{2}\theta}{(3+p^{2}cos2\theta)}\,.  \] 
Solving for  the polynomial equation (6) for $t^{1}$, $t^{2}$, we obtain $\hat{Q}_{1}$ = $\hat {z}_{0}$\,, $\hat{Q}_{2}$ = $\vec{p}(1)$ and 
$\hat{Q}_{3} = \vec{p}(2)$\,.
Thus the invariants associated with the most general mixed spin-1 state are found to be  
\begin{equation} I_{1} = \frac{2\sqrt{6} p|cos\theta|}{(3+p^{2}cos2\theta)}\,\,,\, 
I_{2} = \frac{2\sqrt{3}p^{2}}{(3+p^{2}cos2\theta)}\,\,,\,I_{3} = I_{4}= cos\theta \,\,, I_{5} = -\frac{cos2\theta}{\sqrt 3}  .\end{equation}
Note that in both pure as well as mixed state, $I_{3} = I_{4} = -\frac {cos\theta}{\sqrt 3} , I_{5} = -\frac {cos2\theta}{\sqrt 3}$. For p = 1 and  
$\theta = 0 , \pi$ , the state is separable as in the case of pure state. For $p < 1$, the state is seperable for a range of  values of $\theta$. 
It is observed that as p decreases, the region of $\theta$ for which entanglement appears also decreases \cite{Sirsi}.
 
\section{Conclusion}
\pst
We have considered symmetric N-qubit density matrix expressed in terms of Fano statistical tensor parameters. 
Making use of the well known multiaxial decomposition of the density matrix we have enumerated SU(2) invariants of the most general symmetric state. 
Considering the special case of two qubit symmetric state we have explicitly computed 5 invariants which form a complete set. Our framework can be 
applied to enumerate a complete set of invariants of any qudit state. The study of the relationship between various measures of entanglement and our 
complete set of invariants is in progress.

\section*{Acknowledgments}
\pst
One of us (VA) acknowledges with thanks the University Grants Commission (UGC), India for financial support 
through Faculty Development Programme (FDP).

\end{document}